\documentclass[showpacs,twocolumn,floatfix,superscriptaddress]{revtex4-1}

\usepackage[utf8]{inputenc}
\usepackage{graphics}
\usepackage{color}
\usepackage{stfloats}
\usepackage{amssymb}
\usepackage{amsmath}
\usepackage{overpic}
\usepackage{booktabs, array, mathptmx, tabularx, booktabs, lipsum, amsmath,multirow}
\usepackage{dcolumn}
\usepackage{epstopdf}
\usepackage{booktabs}
\usepackage{bm}
\usepackage{graphicx}
\usepackage{subfigure}
\usepackage{threeparttable}
\usepackage{multirow}

\usepackage{xr}

\begin{document}

\title{Moir\'e flat bands in twisted 2D hexagonal vdW materials}

\author{Qiaoling Xu}
\affiliation{College of Physics and Electronic Engineering, Center for Computational Sciences, Sichuan Normal University, Chengdu 610068, China}
\affiliation{Songshan-Lake Materials Laboratory, Dongguan, Guangdong 523808, China}

\author{Yuzheng Guo}
\email{yguo@whu.edu.cn}
\affiliation{School of Electrical Engineering and Automation, Wuhan University, Wuhan 430072, China}

\author{Lede Xian}
\email{lede.xian@mpsd.mpg.de}
\affiliation{Songshan-Lake Materials Laboratory, Dongguan, Guangdong 523808, China}
\affiliation{Max  Planck  Institute  for  the  Structure  and  Dynamics  of  Matter and Center Free-Electron Laser Science, Luruper  Chaussee  149,  Hamburg 22761,  Germany}

\date{\today}

\begin{abstract}

Moir\'e superlattices in twisted bilayer graphene (TBG) and its derived 
structures can host exotic correlated quantum phenomena because the narrow moir\'e flat minibands in those systems effectively enhance the electron-electron interaction. Correlated phenomena are also observed in 2H-transitional metal dichalcogenides moir\'e superlattices. However, the number of moir\'e systems that have been explored in experiments are still very limited. Here we theoretically investigate a series of two-dimensional (2D) twisted bilayer hexagonal materials (TBHMs) beyond TBG at fixed angles of 7.34$^\circ$ and 67.34$^\circ$ with 22 2D van der Waals (vdW) layered materials that are commonly studied in experiments. First-principles calculations are employed to systemically study the moir\'e minibands in these systems. We find that flat bands with narrow bandwidth generally exist in these systems. Some of the systems such as twisted bilayer In$_2$Se$_3$, InSe, GaSe, GaS and PtS$_2$ even host ultra-flat bands with bandwidth less than 20 meV even for such large angles, which make them especially appealing for further experimental investigations. We further analysis the characters of moir\'e flat bands and provides guidance for further exploration of 2D moir\'e superlattices that could host strong electron correlations.

\end{abstract}

\maketitle

\section{\textbf{INTRODUCTION}}

Moir\'e superlattice (MSL) is a special type of 2D layered material, generated by stacking 
2D vdW materials with a small lattice mismatch or with a twist angle, including graphene, hexagonal boron nitride (hBN), transition metal dichalcogenides (TMDs), various 2D magnets and superconductors  \cite{andrei_marvels_2021}. Different from their parent 2D materials, MSLs with emerging global symmetry and periodicity exhibit fascinating quantum phenomena due to periodic moir\'e modulation of onsite potentials, interlayer coupling and intralayer atomic strain, such as the formation of second-generation Dirac cones \cite{wang_gaps_2016},
Hofstadter butterfly states \cite{hunt_massive_2013} and shear solitons and topological point defects
\cite{gargiulo_structural_2017,wijk_relaxation_2015,alden_strain_2013}. 

Recent breakthroughs on the discovery of correlated insulator states and superconductivity in TBG \cite{cao_correlated_2018, cao_unconventional_2018}
have inspired intensive research on understanding the electronic structures \cite{po2018origin,yuan2018model,koshino2018maximally, kang2018symmetry, song2019all, po2019faithful, tarnopolsky2019origin, pal2019emergent, zhang2019twisted}, the correlated insulating phase, \cite{isobe2018unconventional, po2018origin, liu2018chiral} and the mechanism of superconductivity
\cite{isobe2018unconventional, po2018origin, liu2018chiral,xu2018topological,wu2018theory, lian2019twisted, roy2019unconventional}.
Moreover, novel quantum phenomena are found in TBG, such as correlation induced orbital magnetism
\cite{liu_orbital_2021, andrei_marvels_2021, lu_superconductors_2019} and quantum anomalous Hall states 
\cite{serlin_intrinsic_2020, wu_collective_2020, liu_theories_2021, shi_moire_2021}, 
cascades of phase transitions\cite{zondiner2020cascade}, Chern insulators
\cite{chen_tunable_2020, abouelkomsan_particle_2020}, 
and unconventional ferroelectricity \cite{zheng_unconventional_2020,yasuda_stacking_2021} etc. The exotic correlated properties of TBG are believed to be related to the emergence of the flat bands and the quenching of kinetic energy scales in those states when the twist angle is around the magic angles \cite{bistritzer2011}.  

 Following the success of TBG, a number of new moir\'e flat-band systems beyond TBG are explored, from the homobilayers, heterobilayers to multilayers configurations, including trilayer Graphene/hBN \cite{chen_signatures_2019,chen_evidence_2019},
twisted double bilayer graphene \cite{liu19,shen2020correlated,cao2020tunable,lee2019theory,tutuc2019,he2020tunable,zhang_visualizing_2021}, twisted trilayer graphene \cite{park2021tunable, zhu2020twisted, morell2013electronic}, twisted monolayer–bilayer graphene
 \cite{park2020gate, chen2021electrically,rademaker2020topological,xu2021tunable} and twisted transition metal dichalcogenides (TMDs) \cite{kang2013electronic,Naik18,Wu17,Wu19,zhang2020moire,regan2020,tang2020} and so on.  For example, Dai et al. \cite{liu2019quantum} theoretically studied stacking configurations by the generic form of (M + N)-layers TBG, where the N-layers graphene are stacked on top of M-layers graphene at a small twist angle and explore their electronic structures and topological properties. Here twisted double bilayer graphene, as one of the simplest example with M = 2, N = 2, has already been reported in experiments \cite{liu19,shen2020correlated,cao2020tunable}.  In view of TMDs, numerous stacking configurations with various combination (i.e, twisted MoS$_2$/MoS$_2$ \cite{xian2021realization,angeli2021gamma}, WSe$_2$/WSe$_2$ \cite{wang2020correlated,zhang2020flat}, WS$_2$/WS$_2$ \cite{liao2020precise}, WS$_2$/WSe$_2$ \cite{yuan2020twist}, MoSe$_2$/WSe$_2$ \cite{seyler2019signatures,tran2019evidence,brotons2020spin,rosenberger2020twist,bai2020excitons}, WS$_2$/MoSe$_2$ \cite{zhang2020twist} etc.) are investigated. For instance, the 3D reconstructed WSe$_2$/WS$_2$ moir\'e superlattices with strain are found with the moir\'e flat bands by Crommie et al. recently \cite{li_imaging_2021}. Furthermore, Wang et al. reported a twisted bilayer WSe$_2$ at a small twist angle, where low-energy flat bands are observed and the correlated electronic phases are investigated \cite{wang2020correlated}. 
 
 Except for the typical materials of graphene-based and TMDs, the twisted bilayer black phosphorus \cite{kang2017moire} and grey antimonene \cite{an_moire_2021} are also studied in structural and electronic properties respectively by Guo et al. More recently, Liu et al. investigated antiferro- and ferroelectric bilayer In$_2$Se$_3$ with large twist-angles \cite{li_extremely_2021} based on the first-principles calculations, where low-energy extremely flat band is found. Moreover, the studies of MSLs are further extended to the complex magnetic materials \cite{xiao2021magnetization, hejazi2020noncollinear, wang2020stacking, li2020moire}. Balents et al. theoretically investigated the twisted bilayers of vdW magnets in the structures and phases \cite{hejazi2020noncollinear}, while Tong et al. considered the twisted bilayer 2D magnets CrX$_3$ (X=Br, I) from the magnetization textures aspect \cite{xiao2021magnetization}. Furthermore, other moir\'e dimensionalities are also explored from quasi-one dimension \cite{kennes2020one} up to three dimension \cite{wu2020three,xian2021engineering,song2021eshelby}, which greatly extends the use of twistronic in multi-dimensional systems.
 
 The highly tunable correlation and superconductivity properties of MSLs also make them appealing for future technology applications. A couple of the early attempts were made by Jarillo-Herrero et al. and Rickhaus et al., who made use of the gate-tunable correlated and superconductivity phase of magic angel TBG to fabricate Josephson junctions in single-crystal nanostructures \cite{rodan2021highly, de2021gate}. Moreover, Jarillo-Herrero et al. showed signatures of unconventional ferroelectricity in the bilayer graphene/boron nitride moir\'e system, which may lead to ultrafast, programmable and atomically thin carbon-based memory device applications \cite{zheng2020unconventional}.    

The studies of the MSLs have provided powerful venue to explore the correlated physics  and unconventional superconductivity\cite{kennes_moire_2021,he_moire_2021, andrei_marvels_2021, mcgilly2020visualization} as well as their applications in future technology. In the meanwhile, the emergence of thousands of new vdW layered materials \cite{mounet2018two} gives access to tremendous opportunities for the research of different types of MSLs. However, it remains unclear that whether moir\'e flat bands are generally exist in MSLs when the twist angle is small enough or the moir\'e periodic is large enough. Moreover, it is unclear what types of 2D materials are more susceptible to the formation of moire flat bands and how one could find moir\'e flat band systems with a relatively large twist angle or small system size, which may potentially give rise to stronger electron-electron correlation and/or higher transition temperature for unconventional superconductivity. To address these questions, we perform first-principle calculations to study systematically 22 twisted homo-bilayer superlattices constructed with 2D materials that are accessible in experiments. We constraint our calculations to systems with twist angles at 7.34$^\circ$ and 67.34$^\circ$ as we want to look for moir\'e flat-band systems with small system size. By analyzing the band structures, we find bands with significantly reduced bandwidths generally exist in these MSLs. Interestingly, we do find a few MSLs that can host ultra-flat bands with bandwidth less than 20 meV even for such a large twist angle and small system size. Together with the relatively strong Hubbard interaction, these systems are expected to host strong electron-electron correlations, which is extremely appealing for further experimental investigation. We further discuss the band characters of the parent 2D materials that lead to the formation of the ultra-flat bands and provide guidance for future exploration of numerical exotic strongly-correlated MSLs.    

\section{\textbf{MODEL AND COMPUTATIONAL APPROACHES}}

\begin{figure*}[h]
\centering
\includegraphics [width=6.0in] {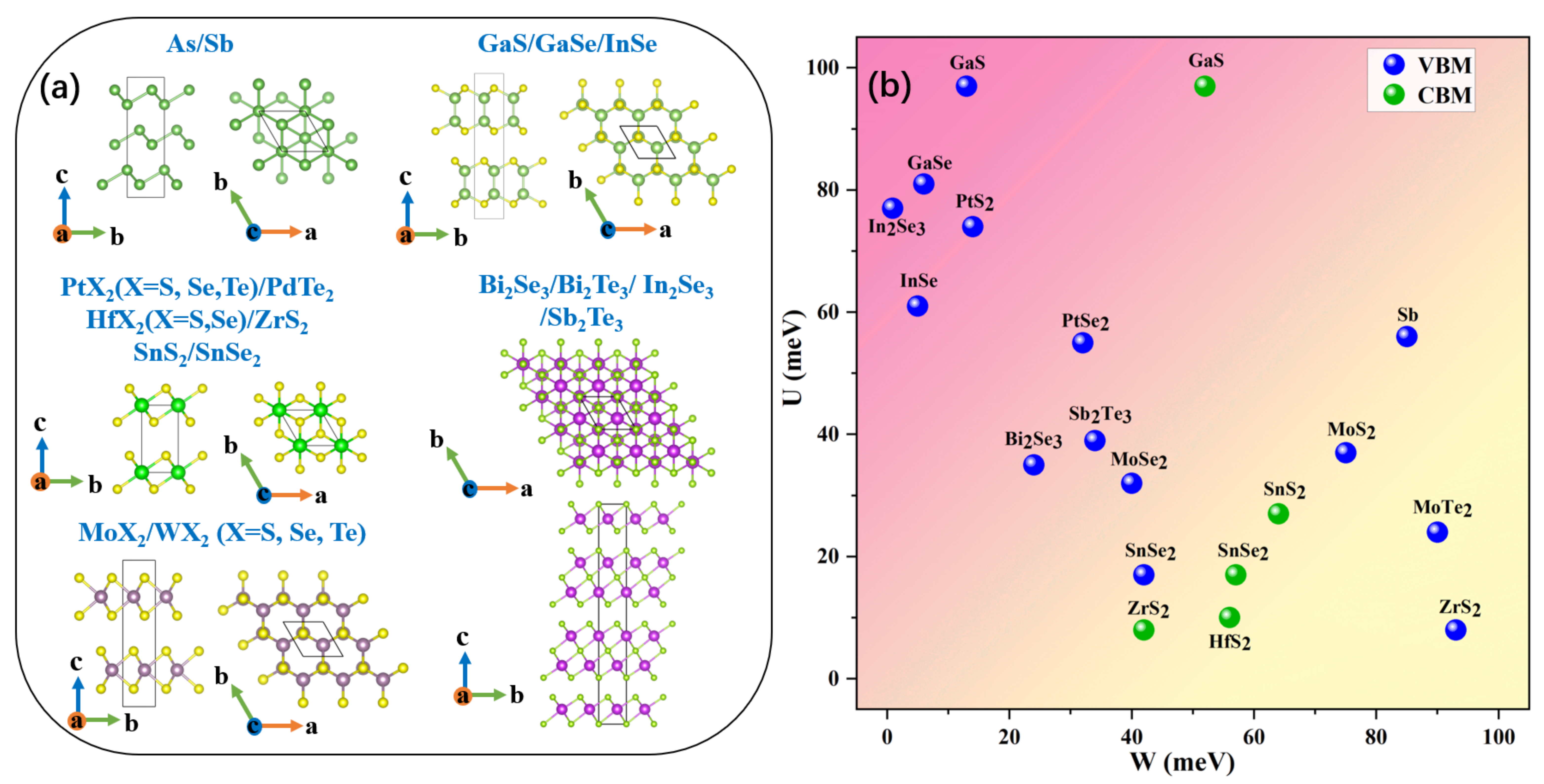}
\caption{(a) The composition and crystal structures of parent 3D vdW materials including
two-atom-thick rhombohedral layered configuration (with the symmetry P63/mmc): As and Sb;
three-atom-thick octahedral layered configuration (with the symmetry P-3m):
PtX$_2$, PdX$_2$, HfX$_2$, ZrX$_2$ and SnX$_2$, or trigonal prismatic coordination
(with the symmetry P63/mmc): MoX$_2$ and WX$_2$;
four-atom-thick layers (with the symmetry R-3mH): GaS, GaSe and InSe;
five-atom-thick layers (with the symmetry R-3mH):
Bi$_2$Se$_3$, Bi$_2$Te$_3$, In$_2$Se$_3$ and Sb$_2$Se$_3$.
(b) Distribution map of the TBHMs onto the variables of
the on-site Hubbard interaction energy U vs bandwidth W
with the scales of no more than 100 meV in VBM and CBM.
The various twisted materials are marked.
The color gradient from the top left (pink) to the bottom right (yellow) of the figure
represents that effect of strong correlation varies from strong to weak.}
\label{Fig1}
\end{figure*}

The present calculations are done within density functional theory (DFT) using the Vienna ab initio software package (VASP) \cite{kresse_1996}. For these configurations of 2D twisted superlattice, the exchange correlation functionals of Perdew Burke and Ernzerhof (PBE) \cite{blochl_1994} are used, in conjunction with Tkatchenko-Scheffler (TS) \cite{tkatchenko_2009,anatole_2010} vdW corrections, which has been shown to give results well consistent with the experimental observations in our previous work on TMD-based MSLs \cite{wang2020correlated}.
An energy cutoff of 400 eV for the plane wave basis sets and the $\Gamma$-centered k-meshes of 1$\times$1$\times$1 are used for geometry optimization and electronic structure calculations
(a 18 $\times$ 18 $\times$ 1 k-grid is used for 1 $\times$ 1 unit cell). A vacuum thickness larger than 15 $\AA$ is used to avoid artificial interactions between periodic slab images.
All atoms are fully relaxed with residual force per atom less than 0.01 eV/$\AA$. Considering the computational cost of superlattice calculations, while the internal atomic positions are fully optimized, the lattice constant for moir\'e supercells is fixed to a value such that it corresponds to the experimental lattice constant for a 1x1 unit cell (See Table I for supercell lattice parameter for each MSL). For all calculations, due to the relativistic effect in heavy elements existing in most systems of TBHMs, the spin-orbit coupling (SOC) effect is considered while results without SOC are also calculated for comparison to estimate the effect of the SOC on the moir\'e flat bands.

For a wide variety of 2D materials,
we concentrate on 22 types that are relatively easy to be synthesized and manipulated in experiments to construct twisted bilayer moir\'e superlattices.
The composition and crystal structures for the corresponding bulk materials are shown in Fig.~(\ref{Fig1}). These 2D materials range from two-atom-thick rhombohedral (grey) layered arsenic (As) and antimony (Sb) with spacegroup P63/mmc,
three-atom-thick layers with octahedral (PtX$_2$, PdX$_2$, HfX$_2$, ZrX$_2$, SnX$_2$) with spacegroup P-3m1
or trigonal prismatic coordination (MoX$_2$, WX$_2$) with spacegroup P63/mmc,
four-atom-thick layers (GaS, GaSe, InSe) with space group R-3mH
to five-atom-thick layers (Bi$_2$Se$_3$, Bi$_2$Te$_3$, In$_2$Se$_3$, Sb$_2$Se$_3$) with spacegroup R-3mH
(The details of structural data are presented in Table I).
Moir\'e superlattices are obtained by rotating two identical layers from these parent 2D materials at a small angle.
We consider TBHM models that have a moir\'e wavelength
$\lambda(\theta_{m})= a/(2\sin(\theta_{m}/2))$\cite{lopes_grap_2007,mele_comm_2010}.
For every commensurate twist angle $\theta_m$, the supercell basis vectors are given by
$t_1=ma_1+(m+1)a_2$ and $t_2=-(m+1)a_1+(2m+1)a_2$,
where $a_{1,2}=(\pm1/2,\sqrt{3}/2)a_0$ are the lattice vectors for the primitive cell of the untwisted system, 
$a_0$ is experimental lattice constant, and $\cos(\theta_m)=(3m^2+3m+1/2)/(3m^2+3m+1)$ with $m$ being an integer. We study TBHM with twist angles at 7.34$^\circ$ and 67.34$^\circ$ that correspond to two distinct configurations when m=4. The largest system we study here has a total number of 610 atoms. A series of optimized twisted configurations can be seen in Figs.~2a, 3(a, e), 4(a, e), 5(a, d) and Figs.~S1-S22(a, b) in the Supporting Information (SI).

\section{\textbf{RESULTS AND DISCUSSION}}

\begin{table*}[h]
\caption{The summary for the calculations of 2D twisted materials including: materials index (i), structural formula, moir\'e supercell lattice constant, average interlayer distance (D), bandwidths at the VBM (W-VBM) without (left) and with SOC (right), bandwidths at the CBM (W-CBM) without (left) and with SOC (right), relative permittivity ($\epsilon$), on-site coulomb repulsion energy (U) and the projected orbitals at the VBM and CBM (only the top contributions listed), and the cohesive energy defined as the energy difference between the total energy of a bilayer with that of two well-separated monolayer in the primitive $1\times1$ cell.}
\centering
\resizebox{\textwidth}{!}{
\begin{tabular}{lccccccccccccc} 
\hline
\multicolumn{1}{l|}{\multirow{2}{*}{index(i)}} & \multicolumn{1}{c|}{\multirow{2}{*}{struct.}} & \multicolumn{1}{l|}{\multirow{2}{*}{lattice const. ($\AA$)}} & \multicolumn{3}{c|}{7.34$^\circ$}   & \multicolumn{3}{c|}{67.34$^\circ$}  & \multicolumn{1}{c|}{\multirow{2}{*}{~ $\epsilon$}} & \multicolumn{1}{l|}{\multirow{2}{*}{U(meV)}} & \multicolumn{2}{c|}{orbital} & \multirow{2}{*}{\begin{tabular}[c]{@{}l@{}}cohesive energy\\~ ~ ~ ~ ~(eV)\end{tabular}}  \\ 
\cline{4-9}\cline{12-13}
\multicolumn{1}{l|}{}                          & \multicolumn{1}{l|}{}                         & \multicolumn{1}{l|}{}                                    & \multicolumn{1}{l|}{Aver. D(A)} & \multicolumn{1}{l|}{W-VBM(meV)} & \multicolumn{1}{l|}{W-CBM(meV)} & \multicolumn{1}{l|}{Aver. D(A)} & \multicolumn{1}{l|}{W-VBM(meV)} & \multicolumn{1}{l|}{W-CBM(meV)} & \multicolumn{1}{l|}{}                                & \multicolumn{1}{l|}{}                        & \multicolumn{1}{c|}{VBM} & \multicolumn{1}{c|}{CBM}          &                                                                                           \\ 
\hline\hline
1     & Bi$_2$Se$_3$\cite{perez_1999}     & 32.295  & 3.185       & 21/24      & n.a.       & 3.169      & 25/36    & n.a. & 12.8\cite{fang_layer_2020}  &35   & Se-p$_z$              & Bi-p$_z$      & 0.23        \\
2     & In$_2$Se$_3$\cite{kupers2018controlled} & 31.24   & 3.233       & 3/7        & 45/46      & 3.195      & 0.8/0.9    & n.a. & 6\cite{wu_dependent_2015}   &77    & Se-p$_z$,p$_x$,p$_y$   & In-$_s$,p$_z$   & 0.25 \\
3     & Sb$_2$Te$_3$\cite{kokh2014}       & 33.354  & 3.343       & 43/34       & n.a.       & 3.340     & 45/57      & n.a. & 11\cite{xu_elec_2019}        &39   & Te-p$_z$              & Sb-p$_z$  & 0.29      \\
4     & Bi$_2$Te$_3$\cite{atuchin2012}    & 34.283  & 3.376       & n.a./18    & n.a.       & 3.368      & n.a./44    & n.a. & n.a.                        & n.a.  & Te-p$_z$              & Bi-p$_z$    & 0.28    \\
5     & GaS\cite{kuhn_1982}              & 28.054   & 3.320       & n.a.       & 52/52      & 3.319      & 13/13      & n.a. & 5.397\cite{errandonea_1999} & 97  & S-p$_z$/Ga-p$_z$       & S-p$_z$/Ga-p$_x$,p$_y$  & 0.23  \\
6     & GaSe\cite{benazeth_1988}         & 29.288   & 3.466       & n.a.       & n.a.       & 3.471      & 5/6        & n.a. & 6.1\cite{errandonea_1999}   & 81  & Se-p$_z$/Ga-p$_z$      & Ga-s/Se-s,p$_z$  & 0.23 \\
7     & InSe\cite{elsayed_2012}          & 31.269   & 3.305       & n.a.       & n.a.       & 3.311      & n.a./5     & n.a. & 7.6\cite{errandonea_1999}   & 61   & Se-p$_z$/In-p$_z$      & In-s/Se-s & 0.24  \\
8     & As\cite{schiferl1969}            & 28.179   & 3.484       & 23/24      & n.a.       & 3.490      & n.a.       & n.a. & n.a.                        & n.a.   & As-p$_z$               & As-p$_z$,s,p$_x$,p$_y$ & 0.15 \\
9     & Sb\cite{li_2008}                 & 33.64    & 3.170       & 64/85      & n.a.       & 3.188      & n.a.       & n.a. & 8\cite{chen_electronic_2019} & 56  & Sb-p$_z$,/Sb-s,p$_x$,p$_y$ & Sb-p$_z$,-s,p$_x$,p$_y$ & 0.31 \\
10    & MoS$_2$\cite{bronsema_1986}      & 24.681   & 3.140       & n.a.       & n.a.       & 3.160      & 75/75      & n.a. & 15.7\cite{laturia_2018}     & 37    & Mo-d$_{z^2}$/S-p$_z$      & Mo-d$_{z^2}$,d$_{x^2-y^2}$ & 0.31 \\
11    & MoSe$_2$\cite{bronsema_1986}     & 25.687   & 3.311       & n.a.       & n.a.       & 3.340      & 40/40      & n.a. & 17.5\cite{laturia_2018}    & 32    & Mo-d$_{z^2}$/Se-p$_z$     & Mo-d$_{x^2-y^2}$,d$_{z^2}$,d$_{xy}$ & 0.31 \\
12    & MoTe$_2$\cite{opalovskij_1965}   & 27.492   & 3.679       & 85/90      & n.a.       & 3.688      & 73/100     & n.a. & 21.7\cite{laturia_2018}    & 24      & Mo-d$_{x^2-y^2}$           & Mo-d$_{x^2-y^2}$,d$_{z^2}$,d$_{xy}$ & 0.34 \\
13    & WS$_2$\cite{schutte_1987}        & 24.627   & 3.225       & n.a.       & n.a.       & 3.251      & n.a.       & n.a. & 14.2\cite{laturia_2018}     & 41    & W-d$_{z^2}$/S-p$_z$       & W-d$_{x^2-y^2}$,d$_{z^2}$,d$_{xy}$  & 0.28 \\
14    & WSe$_2$\cite{kalikhman1983}      & 25.664   & 3.391       & 160/191    & n.a.       & 3.404      & 155/220   & n.a. & 15.7\cite{laturia_2018}   & 36    & W-d$_{x^2-y^2}$/Se-p$_x$,p$_y$  & W-d$_{x^2-y^2}$,d$_{z^2}$,d$_{xy}$ & 0.28 \\
15    & WTe$_2$\cite{yanaki_1973}        & 27.265   & 3.750       & 130/161    & n.a.       & 3.746      & 128/180   & n.a. & n.a.                      & n.a. & W-d$_{x^2-y^2}$    & W-d$_{x^2-y^2}$,d$_{z^2}$,d$_{xy}$ & 0.32\\
16    & PdTe$_2$\cite{kim_1990}          & 31.496   & 2.776       & n.a.       & n.a.       & n.a.       & n.a.       & n.a. & n.a.                      & n.a.  & n.a.                  & n.a.  & 0.53                        \\
17    & PtS$_2$\cite{kliche_1985}        & 27.663   & 3.145       & 15/14      & n.a.       & 3.179      & n.a. & n.a. & 7\cite{du_elastic_2018} & 74  & S-p$_z$/Pt-d$_{z^2}$/S-p$_x$,p$_y$   & Pt-d$_{xy}$,d$_{x^2-y^2}$,d$_{yz}$,d$_{xz}$ & 0.24 \\
18    & PtSe$_2$\cite{kliche_1985}       & 29.108   & 3.138       & 32/32      & n.a.       & 3.139      & 39/48      & n.a. & 9\cite{du_elastic_2018} & 55 & Se-p$_z$/Pt-d$_{z^2}$/Se-p$_x$,p$_y$ & Pt-d$_{xy}$,d$_{x^2-y^2}$,d$_{yz}$,d$_{xz}$ & 0.28\\
19    & PtTe$_2$\cite{kliche_1985}       & 31.436   & 2.96       & n.a.       & n.a.       & n.a.        & n.a.       & n.a. & n.a.                      & n.a. & n.a.                  & n.a.       & 0.43                   \\
20    & HfS$_2$\cite{hodul1984}          & 28.359   & 3.097      & n.a./158   & 55/56      & 3.109       & n.a./160   & 60/60& 51.3\cite{laturia_2018} &10 & S-p$_x$,p$_y$/S-p$_z$& Hf-d$_{z^2}$/S-p$_z$/Hf-d$_{xz}$,d$_{x^2-y^2}$,d$_{xy}$,d$_{yz}$ &0.30\\
21    & HfSe$_2$\cite{yue2015hfse2}    & 29.243   & 3.227      & n.a./276   & 76/76      & 3.224       & n.a./274   & 79/80 & n.a.   & n.a.      & Se-p$_x$,p$_y$ & Hf-d$_{z^2}$/Se-p$_z$/Hf-d$_{xz}$,d$_{x^2-y^2}$,d$_{xy}$,d$_{yz}$ & 0.29 \\
22    & ZrS$_2$\cite{wiedemeier1986}     & 28.609   & 3.030      & n.a./94    & 42/42      & 3.031       & n.a./93    & 46/46  & 61.3\cite{laturia_2018}  &8    & S-p$_x$,p$_y$/S-p$_z$  & Zr-d$_{z^2}$/S-p$_z$/Zr-d$_{yz}$,d$_{xy}$  & 0.34 \\
23    & SnS$_2$\cite{hazen1978}          & 28.413   & 3.349      & n.a.       & 69/69      & 3.342       & n.a.       & 64/64  & 18.8\cite{kumagai_2016}  &27    & S-p$_x$,p$_y$          & Sn-s/S-p$_x$,p$_y$ & 0.19 \\
24    & SnSe$_2$\cite{busch1961}         & 29.764   & 3.396      & n.a./42    & 72/71      & 3.371       & n.a.       & 57/57  & 28\cite{wang_2017}   &17       & Se-p$_x$,p$_y$         & Sn-s/Se-p$_x$,p$_y$,p$_z$ & 0.20\\       
\hline
\end{tabular} }\label{table1}
\end{table*}

\subsection{Flat bands and electronic correlation in MSLs}

We calculate the band structures of the 22 MSLs at the PBE+TS level with and without SOC. The results are summarized in Table~\ref{table1} and the detailed band structures can be found in the SI. Except for the two twisted metallic systems (i.e., twisted bilayer PdTe$_2$ and PtTe$_2$), we find bands with narrow bandwidths commonly appear at the band edges in these MSL systems regardless of the variety of band structures in the original untwisted form. This suggests moir\'e engineering via creating MSL is quite effective in general in creating flat or narrow bands in 2D semiconductor systems. For the metallic system, the bands near the Fermi level are highly entangled such that it is difficult to identify a well-defined flat band (see Fig. S23). We list the bandwidth of the flat bands appearing at the band edges in these systems in Table~\ref{table1}. Not all systems have a value there either because for some systems the bands at the band edges are still quite dispersive (bandwidth W $>$ 200 meV) or there is no well-defined isolated flat bands at the twist angles we study in this work. Furthermore, we calculate the unfolded band of PtSe$_2$ to further elaborate the moir\'e flat band. As shown in Fig. S25 of SI, the unfolded band structure (Fig. S25 (a)) in the twisted bilayer is drastically different from its counterpart (Fig. S25 (b)) in the untwisted structure. In particular, one can visualize a section of isolated flat band appearing at the VBM near the $\Gamma$ point in the Brillouin zone of the primitive cell. This feature corresponds exactly to the flat bands that we calculated in the supercell Brillouin zone. The unfolded band structure reveals that the flat bands are originated from the linear recombination of the VBM states near the $\Gamma$ point. Similar features of the flat bands have been observed in twisted bilayer graphene \cite{utama2021visualization,lisi2021observation} and WS$_2$/WSe$_2$ moir\'e superlattices \cite{stansbury2021visualizing}. From Table~\ref{table1}, it is clear that it is easier to find flat bands at the VBM than at the CBM. For the flat bands at the VBM (CBM), the bandwidth ranges from 0.9 (42) meV to larger than 100 meV in our calculations with SOC.
Nevertheless, it is surprising to find that the isolated band at the band edges in some twisted bilayer systems (such as twisted In$_2$Se$_3$, GaSe, GaS, InSe, PtS$_2$) can be so flat that its bandwidth is less than 20 meV, even the twist angles we study here are relatively large.  

The existence of flat bands in these MSLs indicates the kinetic energy scale of the electron states in these bands is significantly quenched and the electron-electron interaction and correlation may become important. To further evaluate the correlation effects, we estimate the on-site coulomb repulsion energy U in these system as $e^2$/(4$\pi$$\epsilon$${\epsilon_0}$ $a$), where e is electron charge, $\epsilon_0$ is the vacuum permittivity, $\epsilon$ is relative permittivity and $a$ is the effective linear dimension of each site (here we take the length scale of the moir\'e pattern). Though a combination of relative permittivity $\epsilon$ from the data of other literature and the lattice parameter of the moir\'e unite cell a, we estimate U and the values are depicted in Table~\ref{table1}. Then, we compare the energy scale of the bandwidths W in VBM/CBM and estimated Hubbard U of different MSLs by plotting them in Fig.~\ref{Fig1}b (for MSLs with the same materials, only the smallest value of bandwidth is shown). In the region on the top left of Fig.~\ref{Fig1}b (colored in pink), the Hubbard U is larger than the bandwidth W, which indicates correlation effects will be important; while in bottom right region, the bandwidth W is larger and the electron correlation effect will be weak. Fig.~\ref{Fig1}b shows that the twisted compounds of GaS, GaSe, InSe, In$_2$Se$_3$, PtS$_2$ all have narrow bandwidths and relative strong Hubbard interactions and they are expected to host strong electron correlations even at such a large twist angle of $7.34^\circ$. It is noteworthy that twisted arsenene should have a U value near that of antimonene, comparable or even more larger to its relatively small bandwidth of 24 meV, although we couldn't find the value of its $\epsilon$ in literature and didn't list it in the figure. The other systems near the diagonal line, such as VBM of PtSe$_2$, Bi$_2$Se$_3$, Sb$_2$Te$_3$ and MoSe$_2$ and CBM of GaS, ZrS$_2$, HfS$_2$, SnS$_2$ and SnSe$_2$, show comparable U and W values, indicating electron correlations are also important in these systems. For those systems locate at the lower right region in the diagram, such as MoTe$_2$ and ZrS$_2$, the electron correlation may be less important. We need point out that we only compare the relevant energy scale for twisted systems at a relative large twist angle here, the locations of data points in this diagram (i.e., Fig.~1b) will change as the twist angle decreases. The twisted systems appear to have weaker electron correlations here could become strongly correlated systems at smaller angles.

\begin{figure*}[h]
\centering
\includegraphics [width=4in] {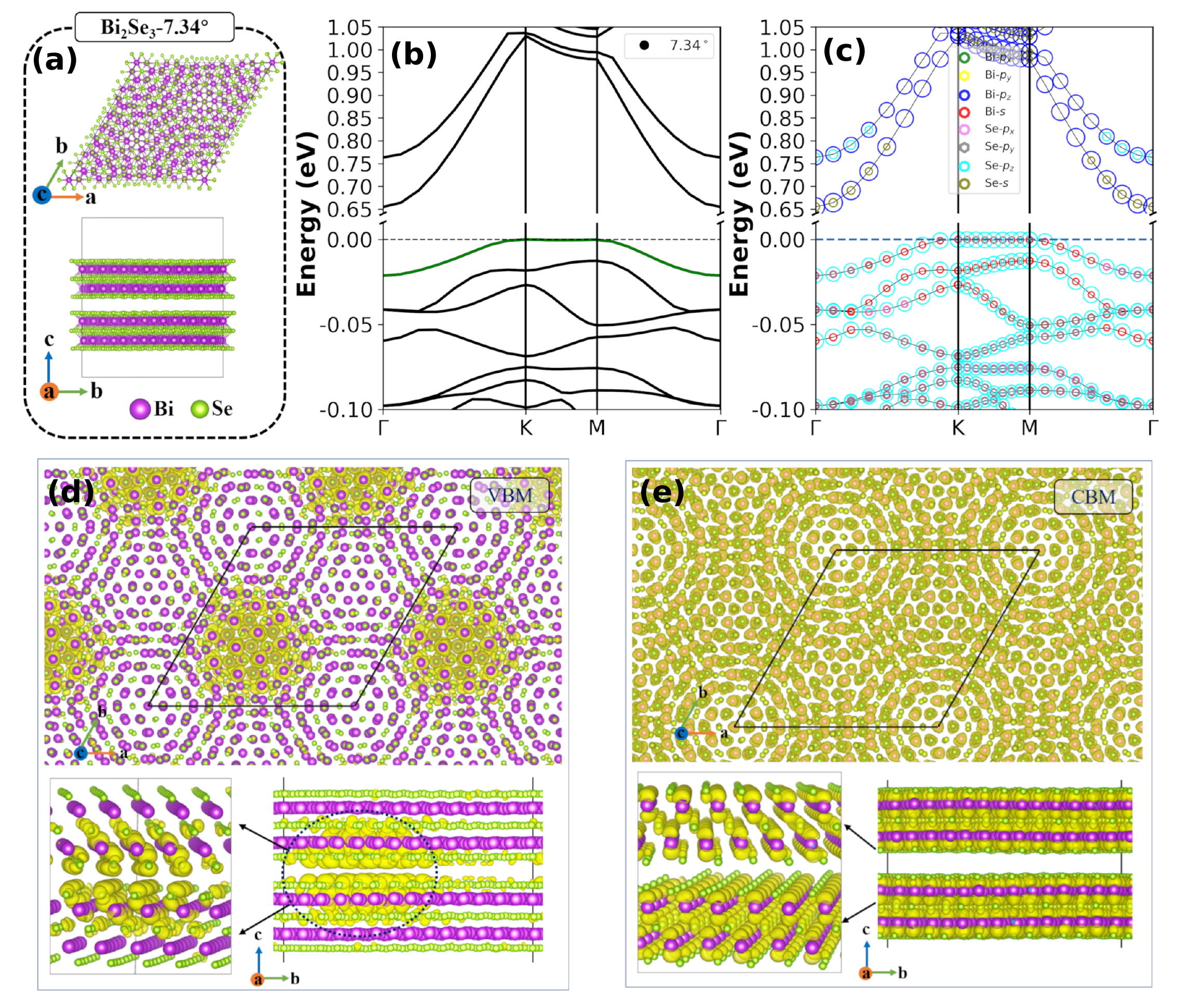}
\caption{Moir\'e flat bands in twisted bilayer Bi$_2$Se$_3$ at 7.34 $^{\circ}$.  (a) Top view (upper panel) and side view (lower panel) of the atomic structure. The purple and green atoms represent Bi and Se atoms, respectively. (b) Low energy band structure near the band edges. (c) Low energy band structure with projection onto each atomic orbital. The size of the circle is proportional to the projection value. 
(d-e) Partial charge density distribution in real space for states at VBM (d) and CBM (e). Upper and lower panel show the top and the side views, respectively. The moir\'e unit cell is indicated by black solid lines. The dashed-line circle in the lower panel of (d) highlights the charge density localization region in Bi$_2$Se$_3$. The isosurface value is set to be 7$\times$$10^{-5}$ $e\AA^{-3}$ }
\label{Fig2}
\end{figure*}

\begin{figure*}[h]
\centering
\includegraphics [width=6in] {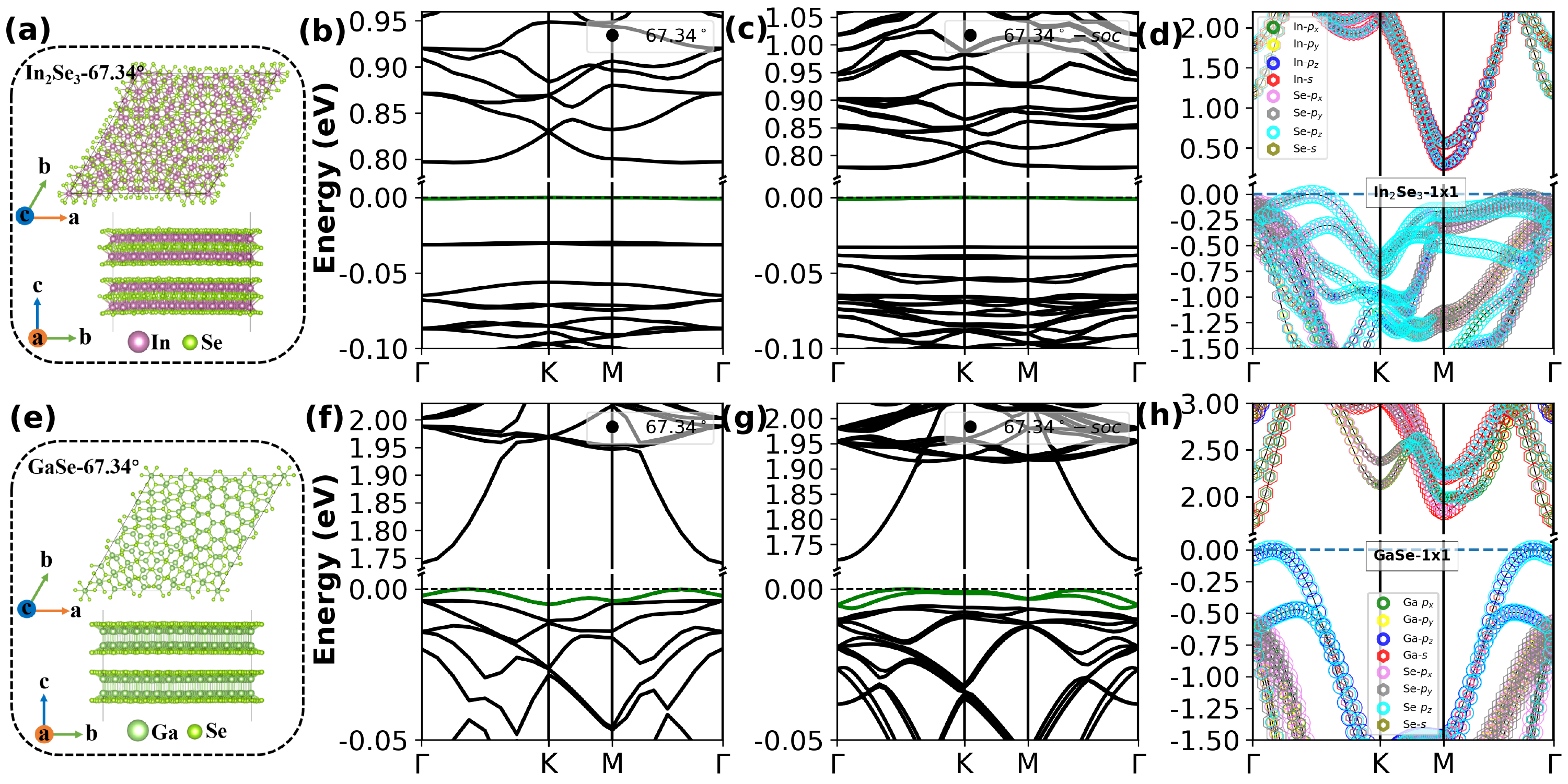}
\caption{(a) Top view (upper panel) and side view (lower panel) of the atomic structure of twisted bilayer In$_2$Se$_3$ at 67.34$^{\circ}$. (b-c) Band structures for twisted bilayer without (b) and with (c) SOC. (d)Projected band structure for untwisted bilayer in 1x1 primitive cell without SOC. (e-h) The corresponding results for the GaSe systems.}
\label{Fig3}
\end{figure*}

\subsection{Characters of the flat bands in MSLs}

To better understand the formation of flat bands in 2D MSLs, we further analysis the characters of the flat bands. To this end, we conduct the calculation of the projected band structures and the results are shown in Figs.~S1-S22 in the SI and the major orbital components for the states in the flat bands are summarized in Table~\ref{table1}. The results reveal that for the flat band systems with smaller dispersion in VBM including the twisted materials of Bi$_2$Se$_3$, In$_2$Se$_3$, Sb$_2$Te$_3$, Bi$_2$Te$_3$, GaS, GaSe, InSe, PtS$_2$ and PtSe$_2$, the valence band edges are associated with the \emph{p$_z$} orbital of the chalcogen atom predominantly. Apart from these materials, the characters are also predominantly \emph{p$_z$} orbitals for both of the twisted arsenene and antimonene in VBM. This is not surprising as we only study MSLs with a relatively large twist angle in this work, the atomic reconstruction has a minor effect. The moir\'e modulation of the band structure is dominated by the modulation of interlayer coupling, instead of the atomic reconstruction as reported in the work by Crommie and others \cite{li_imaging_2021} for MSLs with a much larger system size. Under such circumstance, as the electronic states with p$_z$ orbital are very sensitive to the interlayer coupling, they can be significantly modified by the moir\'e potentials created by the modulated interlayer coupling and turned into flat-band states.    

We take the twisted Bi$_2$Se$_3$ at 7.34$^{\circ}$ (Fig.~2a) as a typical example to further elaborate the role of different atomic orbitals in the formation of flat bands. The calculated band structure and the corresponding projected band structure of twisted Bi$_2$Se$_3$ are given in Fig.~2b and 2c. From these figures, it is clear that the states at the valence band edges are associated with Se \emph{p$_z$} and a fraction of Bi \emph{s} states. Whereas, the states at the conduction band edges are predominantly Bi \emph{p$_z$} states. As the Se \emph{p$_z$} orbitals are located at the outermost Se atomic layer, with charge density extended towards the stacking interface (See lower panels Fig.~2d), they are sensitive to the modulation of the interlayer coupling. Thus those states at the VBM, with relatively large contribution from Se \emph{p$_z$} orbitals, are significantly altered by moir\'e interlayer potentials, forming a flat band with a small bandwidth of 21 meV. The charge density distribution of the states in this flat bands is very localized in the real space as shown in Fig.~2d.  On the other hand, the Bi \emph{p$_z$} orbitals locate at the inner atomic layers and their wavefunction barely extend towards to interface region (see lower panel of Fig.~2e). Therefore, those states at the conduction band edge mainly contributed by the Bi \emph{p$_z$} orbitals are much less sensitive to the modulation of interlayer couplings and the bands at the CBM are still very dispersive. The corresponding charge density distribution of those states is delocalized over the whole moir\'e cell as shown in Fig.~2e. 

Next, we look at those systems that host flat bands with extremely small bandwidths even at a relatively large twist angle. Here, we investigate twisted bilayer In$_2$Se$_3$ and GaSe at 67.34$^{\circ}$ as two typical examples. As shown in Figs.~3b and 3f, the flat bands appear at the VBM with a extremely narrow bandwidth W of 0.8 and 5 meV for twisted bilayer In$_2$Se$_3$ and GaSe at 67.34$^{\circ}$, respectively, in the calculations without SOC. When including SOC, the band structures of both systems have significant modifications: for the In$_2$Se$_3$ system, the band gap between the top and the lower flat bands increases and additional flat bands appear at higher energies; for the GaSe system, a Rashba type of splitting is introduced in the top flat band. Nevertheless, the bandwidths in both systems remain small after including SOC (0.9 meV for the In$_2$Se$_3$ system and 6 meV for the GaSe system). Similar to what we discussed above for the case of twisted Bi$_2$Se$_3$, these ultra-flat bands states are mainly derived from the \emph{p$_z$} orbital of the atoms in the outermost atomic layer (i.e., Se atoms in these cases), as shown in Fig. S24. Moreover, although the chemical composition and atomic structures are different, both systems share similar features in the band structure of the untwisted 1x1 form: that is, a relatively flat plateau at the VBM. This is also the case for the other ultra-flat band systems such as GaS and InSe. Such band plateau, when folded in the moir\'e supercell, naturally appears as flat bands. The moir\'e potential due to the modulated interlayer coupling further introduces band gaps between these flat bands at the supercell Bouillon Zone (BZ) boundary and confines those electronic states, leading to isolated ultra-flat bands. This is actually similar to the trilayer graphene/hBN moir\'e superlattice system \cite{chen_signatures_2019, chen_evidence_2019}, where the states at the band plateau at the VBM are confined by the moir\'e potential formed by the graphene/hBN superlattice. As the band plateau region is more extended in the BZ in these systems, a much smaller moir\'e length scale is sufficient for the formation of ultra-flat bands. It is expected that the moir\'e heterostructures of these 2D layers could also host flat bands at the VBM.   

\begin{figure*}[h]
\centering
\includegraphics [width=6in] {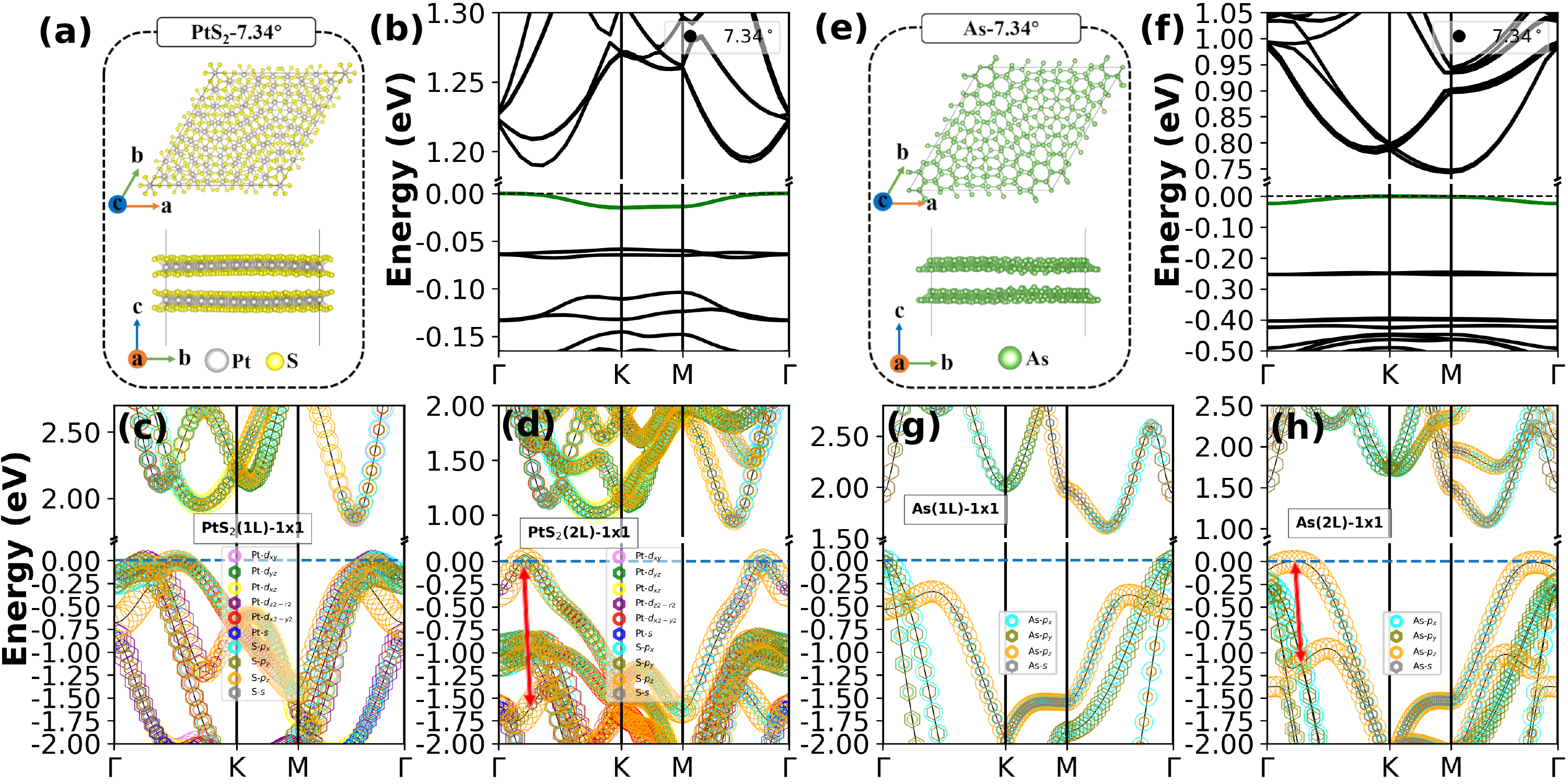}
\caption{ (a) Top view (upper panel) and side view (lower panel) of the atomic structure of twisted bilayer PtS$_2$ at 7.34$^{\circ}$. (b) Band structures for twisted bilayer without SOC. (c-d) Projected band structures for  monolayer (c) and untwisted bilayer (d) in 1x1 primitive cell without SOC. The red arrows highlight the band splitting of the p$_z$ states in the bilayer. (e-h) The corresponding results for the arsenene systems.}
\label{Fig4}
\end{figure*}

\begin{figure*}[h]
\centering
\includegraphics [width=5in] {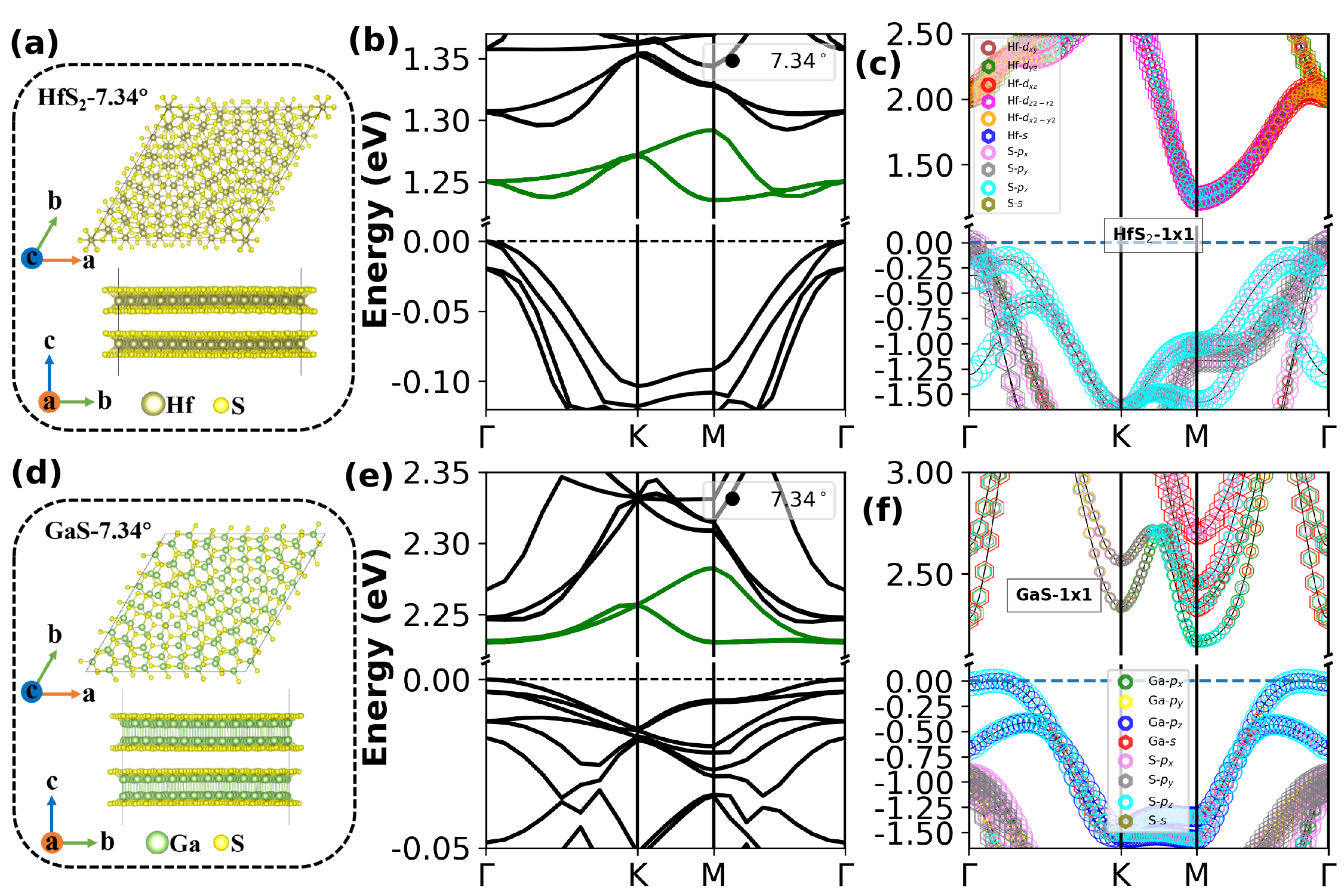}
\caption{(a) Top view (upper panel) and side view (lower panel) of the atomic structure of twisted bilayer HfS$_2$ at 7.34$^{\circ}$. (b) Band structures for twisted bilayer without SOC. (c) Band structure for  untwisted bilayer in 1x1 primitive cell without SOC. (d-f) The corresponding results for the GaS systems.}
\label{Fig5}
\end{figure*}

Twisted bilayer PtS$_2$ is also an interest system. Different from the systems discussed above, although untwisted prinstine bilayer PtS$_2$ does not has a flat band plateau at the band edges, its twisted form at 7.34$^{\circ}$ also host a ultra-flat band with a bandwidth of 15 meV in the calculation without SOC as shown Fig.~4b. The band structure doesn't change much when including SOC (see Fig.~S16 in SI). Although the bandwidth is larger than the cases discussed above, it is still considerable small and comparable to the bandwidth of twist bilayer graphene at the magic angle of 1.05$^{\circ}$, even at such a large twist angle. A noticeable feature in this twisted system is that the separation between flat bands is relatively large, indicating the strength of the interlayer moir\'e potential is relatively large. The relatively large moir\'e potential is likely to be related to the relatively strong interlayer hybridization of the S $p_z$ states. As shown in Fig.~4c and 4d, the S $p_z$ states near the VBM in the untwisted bilayer have a considerable large energy splitting of about 1.5 eV compared with those in the monolayer, which even shifts the VBM from the S $p_x$-$p_y$ states in the monolayer to the S $p_z$ states in the untwisted bilayer. Another system that hosts relatively strong moir\'e potential is twisted bilayer buckled arsenene. As shown in Fig.~4g and 4h, the As $p_z$ states near the VBM in the untwisted bilayer arsenene also have a relatively large energy splitting of about 1.0 eV, shifting the VBM from the As $p_x$-$p_y$ states to the As $p_z$ states. The twisted bilayer arsenene also hosts a large band gap between flat bands as shown in Fig.~4f.

Finally, we discuss the flat bands occurred at the CBM. Generally it is much harder to form flat bands at the CBM. Nevertheless, we found a few exceptions, such as twisted bilayer GaS, 1T-ZrS$_2$, HfS$_2$ and SnS$_2$. The band structures of twisted bilayer 1T-HfS$_2$ and GaS at 7.34$^{\circ}$ without including SOC effect are shown in Figs.~5b and 5e, respectively. The figures clearly show that isolated flat bands with similar shape appear at the bottom of the conduction bands in the two systems, with a bandwidth of 55 (52) meV for 1T-HfS$_2$ (GaS). The flat bands don't change much when the SOC effect is included (see Figs.~S18 and S5 in the SI).  A similar feature for these systems is that the bottom of the conduction bands of the pristine 1x1 bilayer locates at the M point in the BZ (see Fig.~5c and 5f. So the flat bands at the CBM of these systems are formed by the M point states. On the other hand, when the bottom of the conduction bands locate at the Gamma point, the bands at the CBM remain highly dispersive at large twist angles (see the previously discussed twisted bilayer Bi$_2$Se$_3$ and GaSe for example.).

\section{\textbf{Conclusions}}

To summarize, we report a systematic investigation of band structures to explore flat bands in a series of 2D TBHMs with van der Waals beyond TBG using first-principles methods. Two configurations of moir\'e superlattice sharing supercells with the same size at twisted angles of 7.34$^\circ$ and 67.34$^\circ$ are considered. Our calculations show that the twisted compounds of Bi$_2$Se$_3$, In$_2$Se$_3$, Sb$_2$Te$_3$, Bi$_2$Te$_3$, GaS, GaSe, InSe, PtS$_2$, PtSe$_2$, arsenene and antimonene host flat bands at the VBM with the small bandwidths (less than 100 meV). Meanwhile, flat bands also emerge at the CBM for the twisted bilayer materials of GaS, HfS$_2$, HfSe$_2$, ZrS$_2$, SnS$_2$ and SnSe$_2$. Those systems that host flat bands at relatively large twist angles generally have the following characters: (1) The states at the band edges are mainly contributed by the outermost atoms of the layered materials and from those atomic orbitals where the charge density extends towards the stacking interface, such as \emph{$p_z$} and \emph{$d_{z^2}$} orbitals; (2) the band curvatures at around the band edges are relatively large or even relatively flat band plateaus appearing at the band edge in the Brillouin zone of the primitive cell. Then, by estimating the Hubbard interaction U, we find that the twisted compounds of Bi$_2$Se$_3$, In$_2$Se$_3$, GaS, GaSe, InSe and PtS$_2$ exhibit large Hubbard U over bandwidth W ratios, indicating electron correlation may be relatively strong in those systems. Note that our study is limited to the systems with relatively large twist angles. When the twist angle further decreases and the system size becomes larger, flat bands may also appear in other systems, although for those systems the bands at the band edges are still quite dispersive at the angles we studied here. Nevertheless, the ultra-flat band systems we discuss here provide ideal candidates for the study of strong correlation effects at large twist angles. Finally, we discuss the characters of systems that host flat bands at large angles and  provide a guideline for future exploration of novel 2D MSLs that host flat bands and potential strong electron correlations.

\section{\textbf{ACKNOWLEDGEMENT}}
This work is supported by the Key-Area Research and Development Program of Guangdong Province of China
(Grants No.2020B0101340001). The computational resource is provided by the Platform for Data-Driven
Computational Materials Discovery of the Songshan Lake laboratory. 

\bibliography{ref}

\end{document}